\documentclass[onecolumn,showpacs]{revtex4}
\usepackage{amsmath}
\usepackage{amssymb}
\usepackage[dvips]{graphicx}

\begin{document}

\title{ Polaronic memristor strongly coupled to electrodes}
\author{A. S. Alexandrov$^{1}$ and A. M. Bratkovsky$^{2}$}
\affiliation{$^{1}$Department of Physics, Loughborough University, Loughborough LE11 3TU,
United Kingdom\\
$^{2}$Hewlett-Packard Laboratories, 1501 Page Mill Road, Palo Alto,
California 94304\\
}

\begin{abstract}
Attractive electron correlations due to an electron-vibron interaction (EVI)
can overcome the direct Coulomb repulsion of polarons in strongly deformable
molecular quantum dots (MQDs). If it realizes, a switching appears in the
I-V characteristics of the degenerate nonadiabatic molecular bridges weakly
coupled to electrodes providing a route to ultrafast `memristors'
(memory-resistors) as the basis for future oscillators, amplifiers, and
other important circuit elements. Here, we extend our theory of polaronic
memristors to adiabatic MQDs strongly coupled to the leads and show that the
degeneracy of MQD (or other multilevel energy structure) along with the
polaron-polaron attraction is a necessary ingredient of its switching
behavior in the strong-coupling regime as well.
\end{abstract}

\pacs{63.22.Gh, 62.23.Hj, 73.23.Hk, 71.38.Mx}
\maketitle

Different nano-size devices are being proposed and investigated \cite%
{rat,col,rat2} that exhibit some kind of current `switching' behavior \cite%
{don,ste}, `negative differential resistance', and `memory' \cite{was}.
There is currently a surge of interest in various systems showing memristor
behavior (see, for example \cite{stan} and references therein) that can
potentially be used for Resistive Random Access Memories (RRAMs). The
practical significance of determining a precise microscopic mechanism of
such a behavior is difficult to overestimate.

Polarons - electrons strongly coupled with lattice vibrations (phonons or
vibrons) - play a key role in the transport and optical properties of many
systems of reduced dimensions and dimensionality \cite{dev,korkin,bra}. They
may provide an almost instantaneous switching mechanism \cite{alebra}, when
a bistable current state of vibrating nano-circuits appears due to the
attractive electron-electron correlations (`anti-Coulomb blockade' of the
insulating state \cite{alebrastan,erm}), if MQD is many-fold degenerate. A
mean-field approximation with respect to EVI \cite{gal} fails to describe
intrinsic molecular switching, so that the exact analytical or numerical
solutions of a polaronic transport problem are required \cite{alebra,alebra2}%
.

An exact analytical solution of the problem fully accounting for the strong
EVI and polaron-polaron correlations in MQD has been found by us in a weak
molecular-lead coupling limit, where an inverse lifetime, $\Gamma /\hbar $,
of a polaron on the dot is small compared with the characteristic vibron
frequency, $\Gamma /\hbar \ll \omega _{0}$ \cite{alebra} (nonadiabatic
regime). Here we extend our theory of polaronic memristors to MQDs strongly
coupled with the leads, where $\Gamma /\hbar \gg \omega _{0}$ (adiabatic
regime).

The molecular Hamiltonian includes the Coulomb repulsion, $U^{C}$, and the
electron-vibron interaction as \cite{alebra}
\begin{align}
& H_{m}=\sum_{\mu }\varepsilon _{\mu }\hat{n}_{\mu }+\frac{1}{2}\sum_{\mu
\neq \mu ^{\prime }}U_{\mu \mu ^{\prime }}^{C}\hat{n}_{\mu }\hat{n}_{\mu
^{\prime }}  \notag \\
& +\sum_{\mu ,q}\omega _{q}\hat{n}_{\mu }(\gamma _{\mu
q}d_{q}+H.c.)+\sum_{q}\omega _{q}(d_{q}^{\dagger }d_{q}+1/2).  \label{hamm}
\end{align}%
Here $\varepsilon _{\mu }$ are energy levels of a rigid molecule, $\hat{n}%
_{\mu }=c_{\mu }^{\dagger }c_{\mu }$ is the molecular occupation number
operator, $d_{q}$ annihilates vibron, $\omega _{q}$ is the vibron frequency,
and $\gamma _{\mu q}$ is the EVI constant ($q$ enumerates the vibron modes,
and we have taken $\hbar =k_{B}=1$ here and below). This Hamiltonian
conserves the occupation numbers of molecular states $n_{\mu }$. When the
molecule is attached to two (for simplicity symmetric) leads, the complete
Hamiltonian $H=H_{m}+H_{l}+\sum_{\mu ,k}t_{k\mu }c_{\mu }^{\dagger
}(a_{k}+b_{k})+H.c.$ includes the hopping to the leads, and the Hamiltonian
of the leads $H_{l}=\sum_{k}\xi _{k}(a_{k}^{\dagger }a_{k}+b_{k}^{\dagger
}b_{k})$, where $a_{k}$ and $b_{k}$ annihilate the electron in the left and
right electrodes, respectively.

With some conventional assumptions (energy independent width $\Gamma =2\pi
\sum_{k}t_{k\mu }^{2}\delta (E-\xi _{k})$ and quasi-equilibrium electron
distributions in the leads ) the current through MQD is given by the
Landauer-type expression (or by the Fermi golden rule) as \cite{alebra}:
\begin{equation}
I(V)=I_{0}\int_{-\infty }^{\infty }dE\left[ f_{1}(E)-f_{2}(E)\right] \rho
(E),  \label{eq:Is}
\end{equation}%
allowing for a transparent analysis of the essential physics of the
switching phenomenon. Here $I_{0}=e\Gamma $, $f_{1(2)}(E)=1/[\exp [\beta
(E+\Delta \mp eV/2)/T]+1]$ is the electron distribution function in left (1)
and right (2) metallic leads, respectively, $\beta =1/T$ the inverse
temperature, $\Delta $ the position of the lowest unoccupied molecular level
with respect to the Fermi level at $V=0$, and $V$ the voltage drop across
MQD.

The molecular density-of-states (DOS) $\rho (E)=-(1/\pi )\sum_{\mu }\rm{Im}%
G_{\mu }^{R}(E)$ generally depends on all interactions and molecule-lead
hopping integrals as determined by the Fourier component $G_{\mu }^{R}(E)$
of the retarded Green's function (GF) $G_{\mu }^{R}(t)=-i\Theta (t)\left<c_{\mu
}(t)c_{\mu }^{\dagger }+c_{\mu }^{\dagger }c_{\mu }(t)\right>,$ where $c_{\mu }(t)$
is the Heisenberg annihilation operator. For calculating $\rho (E),$ one can
neglect molecule-lead coupling in the weak-coupling regime while keeping all
orders of the Coulomb repulsion and EVI by means of the canonical
displacement transformation of the molecular Hamiltonian (\ref{hamm}). In
particular, assuming EVI with a single vibronic mode, $\omega _{q}=\omega
_{0}$ and the coupling $\gamma _{\mu q}=\gamma $, and constant Coulomb
integrals, $U_{\mu \mu ^{\prime }}^{C}=V_{c}$, one obtains for a $d$-fold
degenerate single-level dot \cite{alebra},
\begin{eqnarray}
&&\rho (E)=\mathcal{Z}d\sum_{r=0}^{d-1}Z_{r}(n)\sum_{l=0}^{\infty
}I_{l}\left( \xi \right)   \notag \\
&&\times \biggl[e^{\frac{\beta \omega _{0}l}{2}}\left[ (1-n)\delta
(E-rU-l\omega _{0})+n\delta (E-rU+l\omega _{0})\right]   \notag \\
&&+(1-\delta _{l0})e^{-\frac{\beta \omega _{0}l}{2}}[n\delta (E-rU-l\omega
_{0})+(1-n)\delta (E-rU+l\omega _{0})]\biggr].  \label{eq:rho}
\end{eqnarray}%
Here $\mathcal{Z}=\exp \left[ -|\gamma |^{2}\coth (\beta \hbar \omega _{0}/2)%
\right] $ accounts for a familiar polaronic renormalization of the hopping
integrals, $\xi =|\gamma |^{2}/\sinh (\beta \hbar \omega _{0}/2),$ $%
I_{l}\left( \xi \right) $ is the modified Bessel function, and $\delta _{lk}$
is the Kroneker symbol ($l,k=0,1,2,...)$, and we have taken the position of
the level as zero, $\epsilon _{\mu }=0$. The resulting polaron-polaron
interaction, $U=V_{c}-2E_{p}$, where $E_{p}=|\gamma |^{2}\omega _{0}$ is the
polaron level shift, comprises the Coulomb repulsion, $V_{c}$, and the
vibron-mediated attraction. An important feature of DOS (\ref{eq:rho}) is
its nonlinear dependence on the occupation number $n$ of degenerate
molecular states owing to the \emph{correlation }side-bands with the spectral
weight $Z_{r}(n)=(d-1)!n^{r}(1-n)^{d-1-r}/(r!(d-1-r)!).$ The DOS, Eq.(\ref%
{eq:rho}), contains full information about all possible correlations in
transport, in particular, the vibron and correlation side-bands. It is
derived by solving the finite system of coupled equations for $N$-particle
Green's functions, as described in Ref. \cite{alebra}.

Equating incoming and outgoing numbers of electrons in MQD per unit time, one
obtains the self-consistent equation for the molecular state occupation
number $n$ as \cite{alebra}:
\begin{equation}
2nd=\int dE\rho \left( E\right) \left[ f_{1}(E)+f_{2}(E)\right] ,  \label{n}
\end{equation}%
which automatically satisfies the condition $0\leqslant n\leqslant 1$.
\begin{figure}[tbp]
\begin{center}
\includegraphics[angle=0, width=0.4\textwidth]
{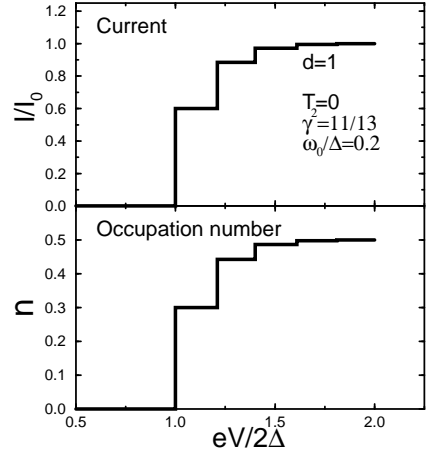}
\end{center}
\caption{Current-voltage characteristic of the nondegenerate ($d=1)$ MQD
weakly coupled to the leads at $T=0$K, $\protect\omega _{0}/\Delta =0.2$,
and $\protect\gamma ^{2}=11/13$. The vibron ladder in clearly seen in the I-V characteristic, 
but there is no hysteresis.}
\label{fig:deg1}
\end{figure}
In particular, for the \emph{nondegenerate} MQD ($d=1$) and $T=0$K the
result is
\begin{equation}
n=\frac{b_{0}^{+}}{2+b_{0}^{+}-a_{0}^{+}},
\end{equation}%
and
\begin{equation}
{\frac{I}{{I_{0}}}}=\frac{2b_{0}^{-}+a_{0}^{-}b_{0}^{+}-a_{0}^{+}b_{0}^{-}}{%
2+b_{0}^{+}-a_{0}^{+}}.
\end{equation}%
where $a_{0}^{\pm }=\mathcal{Z}\sum_{l=0}^{\infty }(|\gamma
|^{2l}/l!)[\Theta (l\omega _{0}-\Delta +eV/2)\pm \Theta (l\omega _{0}-\Delta
-eV/2)],$ and $b_{0}^{\pm }=\mathcal{Z}\sum_{l=0}^{\infty }(|\gamma
|^{2l}/l!)[\Theta (-l\omega _{0}-\Delta +eV/2)\pm \Theta (-l\omega
_{0}-\Delta -eV/2)]$. Here $\Theta (x)=1$ if $x>0$ and zero otherwise. The
current is single-valued, Fig.~\ref{fig:deg1}, with the familiar ladder due
to the phonon side-bands. The double-degenerate level $d=2$ obviously
provides greater number of elementary processes for conductance reflected in
a larger number of steps on the phonon ladder in comparison with $d=1$ case.
However, the current remains single-valued showing no switching behavior
\cite{alebra,bra} in the weak molecular-lead coupling regime, contrary to
the mean-field approximation (MFA) of Ref. \cite{gal}, which yields an
artificial switching behavior for a non-degenerate level, $d=1$, in this
regime. In fact, the switching appears only in the I-V curves of many-fold
degenerate dot when $U<0$ and the temperature is low enough \cite{alebrastan}%
.

We analyze below an effect of a strong molecular-lead coupling $\Gamma \gg
\omega _{0}$ on these results. When EVI is also large ($\gamma \gg 1$ ), it
could not be treated perturbatively, so that one has to sum all diagrams of
the perturbation expansion in powers of $\gamma $ in order to calculate GFs.
As suggested in Ref. \cite{gal2}, this can be done with a lowest-order
non-equilibrium linked cluster expansion (NLCE) \cite{kral}, which provides
an approximate resummation of the whole series for GFs as:
\begin{equation}
G(t)\equiv \sum_{r=0}^{\infty }\gamma ^{2r}W_{r}(t)\approx G_{0}(t)\exp
[\gamma ^{2}F(t)].  \label{nlce}
\end{equation}%
Here $G_{0}(t)$ should be calculated without EVI [i.e. with $\gamma _{\mu
q}=0$ in Eq.(\ref{hamm})], while $F(t)$ is found using the lowest
(second)-order diagrams $F(t)=G_{0}^{-1}(t)W_{1}(t)$.

\begin{figure}[tbp]
\begin{center}
\includegraphics[angle=-90, width=1.0\textwidth]
{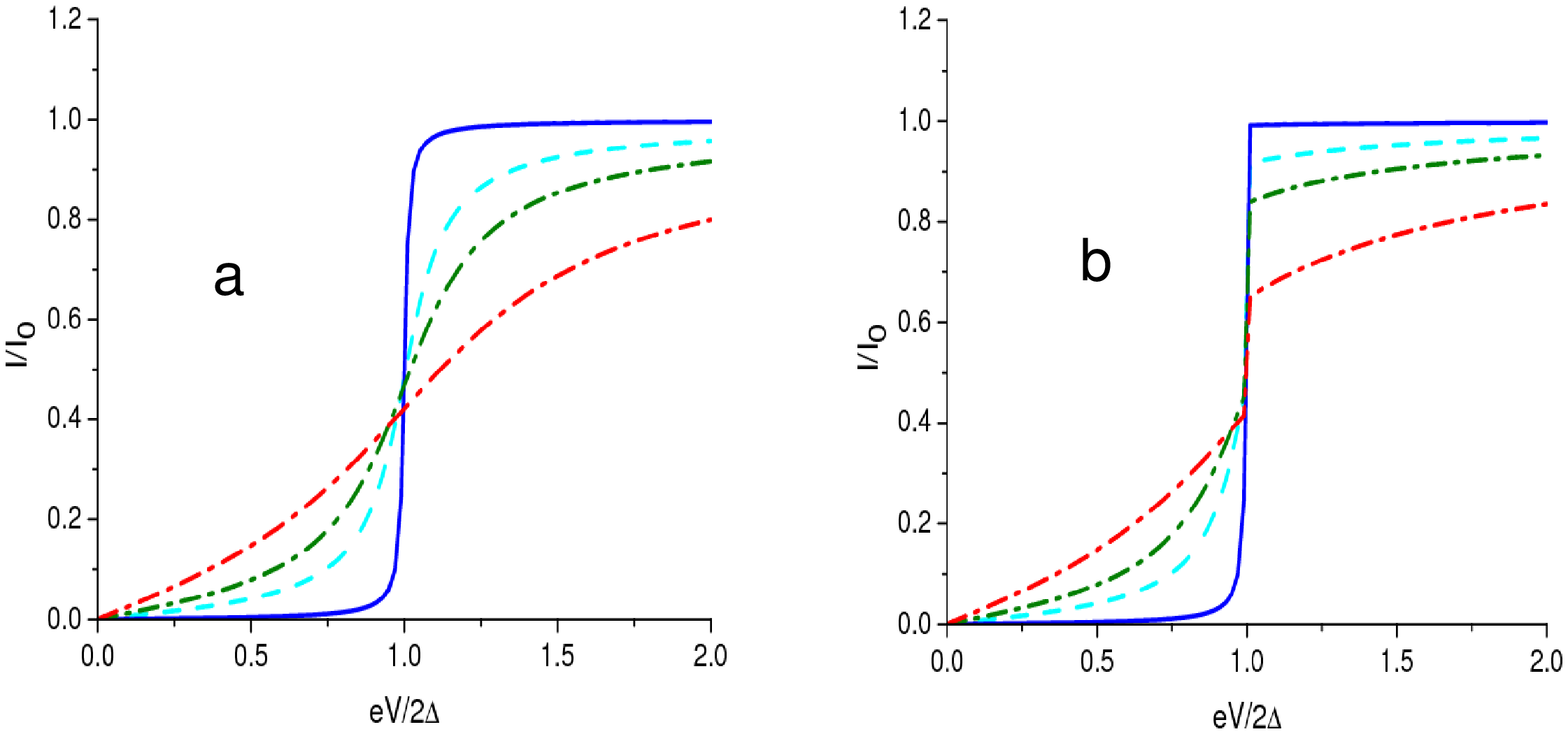}
\end{center}
\caption{Current-voltage characteristics of the nondegenerate ($d=1)$ rigid
molecular bridge at $T=0$K ($a$), and of the non-degenerate adiabatic ($%
\protect\omega _{0}\ll \Gamma $)  molecular bridge with the polaron level
shift $E_{p}=\Delta /2$ ($b$) for different couplings to the leads, $\Gamma
/\Delta =0.01$ (the steepest curve), $0.1,$ $0.2,$ and $0.5.$}
\label{fig2}
\end{figure}

Calculating zero $G_{0}(t)$ and second order (in $\gamma $) $W_{1}(t)$ is a
textbook exercise for the non-degenerate molecular level. Using equations of
motion, $i\dot{c}=t\sum_{k}(a_{k}+b_{k})$ and $i\dot{a}_{k}=\xi _{k}a_{k}+tc$%
, $i\dot{b}_{k}=\xi _{k}b_{k}+tc$, one can readily find the Fourier component
of the retarded GF of the rigid non-degenerate dot as $G_{0}^{R}(E)=(E-i%
\Gamma )^{-1}$, and a lorentzian zero-order molecular DOS,
\begin{equation}
\rho _{0}(E)={\frac{1}{{\pi }}}{\frac{\Gamma }{{E^{2}+\Gamma ^{2}}}}.
\label{dos2}
\end{equation}%
Substituting Eq.(\ref{dos2}) into Eq.(\ref{eq:Is}) yields the current
through a rigid bridge as
\begin{equation}
{\frac{I^{(0)}}{{I_{0}}}}={\frac{1}{{\pi }}}\left[ \arctan \left( {\frac{%
eV/2-\Delta }{{\Gamma }}}\right) +\arctan \left( {\frac{eV/2+\Delta }{{%
\Gamma }}}\right) \right] ,
\end{equation}%
shown in Fig.~\ref{fig2}$a$ at $T=0$K for a few inverse lifetimes.

Using the zero-order lesser, $G_{0}^{<}(t)\approx in_{0}\exp (-\Gamma |t|)$,
and greater, $G_{0}^{>}(t)\approx -i(1-n_{0})\exp (-\Gamma |t|)$, GFs \cite%
{gal2} one can readily calculate the second order contribution $W_{1}(t)$.
Here $n_{0}=[f_{1}(0)+f_{2}(0)]/2$ is the \emph{zero-order} population of
the molecular level. Then applying NLCE, Eq.(\ref{nlce}), one gets the
retarded GF for the strong molecular-lead coupling, $\Gamma \gg \omega _{0}$%
, summing infinite number of EVI diagrams (see for details Ref. \cite%
{gal2,kral}), $G^{R}(E)=(E+2n_{0}E_{p}-i\Gamma )^{-1}$, and the molecular
DOS,
\begin{equation}
\rho (E)={\frac{1}{{\pi }}}{\frac{\Gamma }{{(E+2n_{0}E_{p})^{2}+\Gamma ^{2}}}%
}.  \label{dos3}
\end{equation}%
Using Eq.(\ref{dos3}) in Eqs.(\ref{n}),(\ref{eq:Is}), one easily obtains:%
\begin{equation}
n={\frac{1}{{2\pi }}}\left[ \pi +\arctan \left( {\frac{eV/2-\tilde{\Delta}}{{%
\Gamma }}}\right) -\arctan \left( {\frac{eV/2+\tilde{\Delta}}{{\Gamma }}}%
\right) \right] ,  \label{n2}
\end{equation}%
for the molecular-state population and
\begin{equation}
{\frac{I}{{I_{0}}}}={\frac{1}{{\pi }}}\left[ \arctan \left( {\frac{eV/2-%
\tilde{\Delta}}{{\Gamma }}}\right) +\arctan \left( {\frac{eV/2+\tilde{\Delta}%
}{{\Gamma }}}\right) \right] ,  \label{I2}
\end{equation}%
for the current, respectively, with $\tilde{\Delta}\equiv \Delta
-2n_{0}E_{p}=\Delta -E_{p}\Theta (eV/2-\Delta )$ at low temperatures, $T\ll
\Gamma $. The level population, Eq.(\ref{n2}), of the non-degenerate dot
strongly coupled to the leads, and the current, Eq.(\ref{I2}), Fig.~\ref%
{fig2}b, remain single-valued showing neither switching nor negative
differential resistance similar to the weak molecular-lead coupling regime,
Fig.~\ref{fig:deg1}. Taking into account the coupling with the leads beyond
the second order washes out the vibron ladder in the I-V characteristic, so that the EVI
affects the I-V characteristics only marginally when compared with the rigid
dot, Fig.~\ref{fig2}, in the strong-coupling (adiabatic) regime.

As we have shown earlier \cite{alebra2}, the mean-field approximation (MFA)
of Ref. \cite{gal} replacing the occupation number operator $\hat{n}$ in EVI
by the average population $n$ erroneously leads to a non-linear equation for
$n$ and a spurious switching of the non-degenerate dot, which has no
physical meaning. The authors of Ref.~\cite{gal2} have confirmed our
conclusion for the weak molecular-lead coupling, but argued by using NLCE
that, surprisingly, their MFA works well in the strong coupling regime, so
that the non-degenerate dot is multi-stable contrary to our present result,
Fig.~\ref{fig2}b. The discrepancy originates in an erroneous replacement of
the zero-order occupation $n_{0}$ in $G_{0}^{<,>}(t)$, $G^{R}(E),$ and as
the result in Eq.(\ref{dos3}), by the \emph{exact} $n$ leading to
double-counting of EVI diagrams in Ref. \cite{gal2}. In fact, multiple
solutions for the steady-state population of the non-degenerate dot found in
Ref. \cite{gal} and Ref.\cite{gal2} are artefacts of MFA and mis-application
of NLCE, respectively. The physical mechanism of the switching is provided
by an attraction of two polarons on a multiply-degenerate MQD \cite%
{alebrastan,alebra,erm}, which is missing in the non-degenerate `polaron
model' of Refs. \cite{gal,gal2} due to the Pauli exclusion principle.

\begin{figure}[tbp]
\begin{center}
\includegraphics[angle=-90, width=1.0\textwidth]
{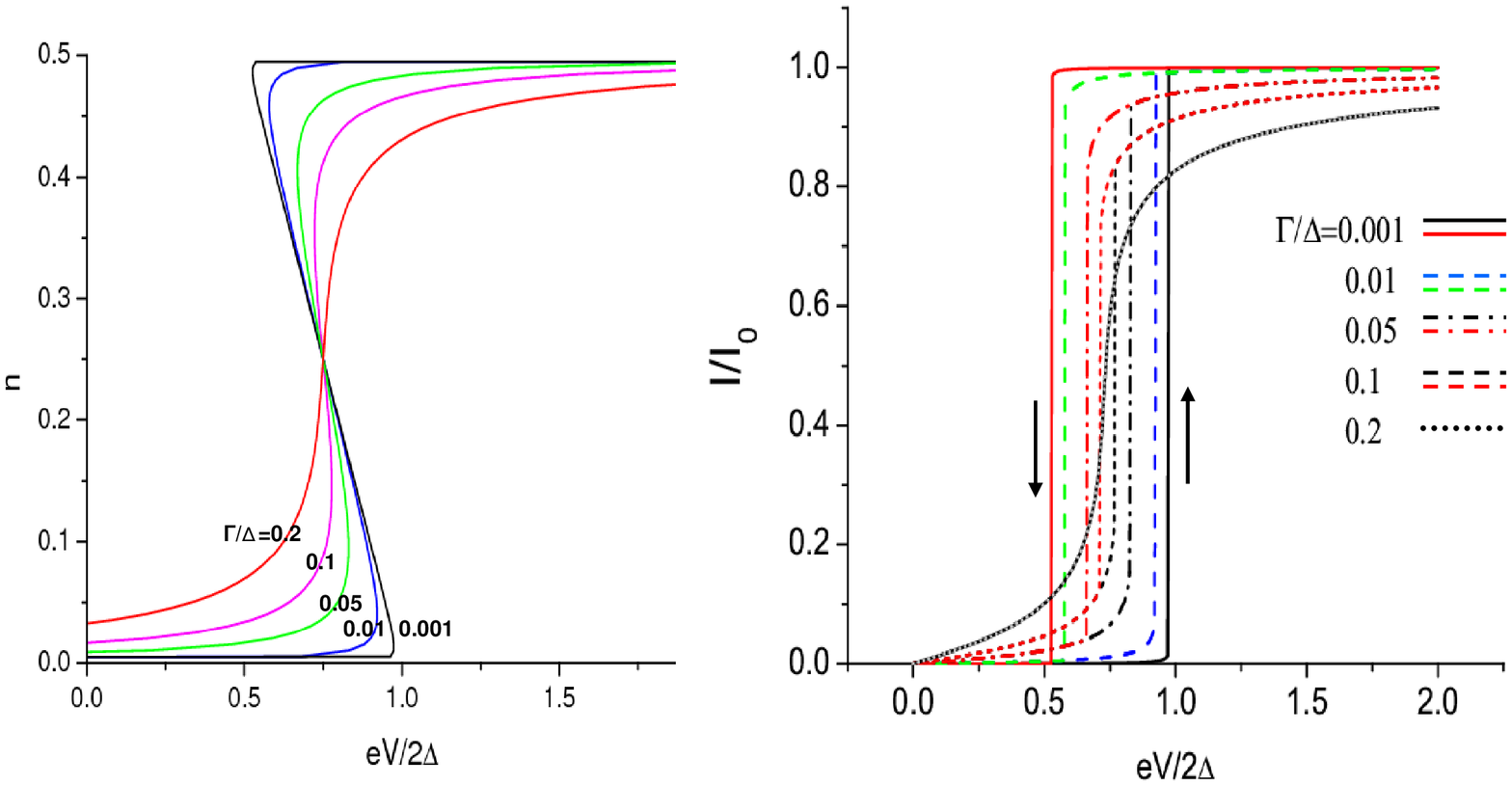}
\end{center}
\caption{Steady-state population $n$ and the current $I/I_{0}$ of the
many-fold degenerate negative$-U$ dot at $T=0$K with different values of the
coupling to the leads, and the effective attraction $\widetilde{U}=-\Delta
/2.$}
\label{fig3}
\end{figure}

Treating the polaron-polaron correlations in real multi-level bridges and
the molecular-lead coupling at the same level of approximation is a
challenging problem, which might require numerical techniques. For instance,
the conductance of deformable molecules has been numerically studied in the
framework of a two-impurity Anderson model with positive and negative
electron-electron interactions and in the two-impurity Anderson-Holstein
model with a single phonon mode \cite{bonca}, where the spin and charge
Kondo effects can occur simultaneously at any coupling strength. Since the
vibron ladders in I-V curves are washed out by the strong coupling with the
leads, Fig.~\ref{fig2}b, we consider here an effect of the molecular-lead
coupling on the switching in the many-fold degenerate ($d\gg 1$) negative
Hubbard $U$ model of Ref.~\cite{alebrastan} allowing for a simple analytical
solution,
\begin{equation}
H_{m}=\frac{1}{2}U\sum_{_{\mu }\neq \mu ^{\prime }}\hat{n}_{_{\mu }}\hat{n}%
_{\mu ^{\prime }}+\sum_{k}\xi _{k}(a_{k}^{\dagger }a_{k}+b_{k}^{\dagger
}b_{k})+\sum_{k,\mu }t_{k\mu }(a_{k}^{\dagger }+b_{k}^{\dagger })c_{\mu
}+H.c.  \label{hub}
\end{equation}%
For the many-fold degenerate dot one can approximate the exact two-body
interaction in the Hamiltonian Eq.(\ref{hub}) by a single-particle
self-consistent Hartree-type potential as $\frac{1}{2}U\sum_{_{\mu }\neq \mu
^{\prime }}\hat{n}_{_{\mu }}\hat{n}_{\mu ^{\prime }}\approx U\sum_{_{\mu
}\neq \mu ^{\prime }}\hat{n}_{_{\mu }}n_{\mu ^{\prime }}-\frac{1}{2}%
U\sum_{_{\mu }\neq \mu ^{\prime }}n_{_{\mu }}n_{\mu ^{\prime }}$. Then,
using equations of motion yields the DOS as in the non-degenerate rigid
dot, Eq.(\ref{dos2}), but shifted in energy by the attraction potential,
\begin{equation}
\rho (E)={\frac{1}{{\pi }}}{\frac{d\Gamma }{{[E-U(d-1)n]^{2}+\Gamma ^{2}}}}.
\label{dos4}
\end{equation}%
From Eq.(\ref{n}), the molecular state population for $T\ll \Gamma $ is:
\begin{equation}
n={\frac{1}{{2\pi }}}\left[ \pi +\arctan \left( {\frac{{eV/2-\Delta -2\tilde{%
U}n}}{{\Gamma }}}\right) -\arctan \left( {\frac{{eV/2+\Delta +2\tilde{U}n}}{{%
\Gamma }}}\right) \right] ,  \label{n4}
\end{equation}%
where $\tilde{U}=U(d-1)/2$.

For $\Gamma =0,$ there are two stable solutions of Eq.(\ref{n4}), $n=0$ and $%
n=0.5$ in the voltage region of bi-stability, $\Delta -|\tilde{U}%
|<eV/2<\Delta $, and only one solution, $n=0.5$ for $eV/2>\Delta ,$ and one
solution $n=0$ for $eV/2<\Delta -|\tilde{U}|$. The coupling with the leads
does not destroy the many-fold degenerate memristor as long as $\Gamma $ is
less than some critical value, $\Gamma <\Gamma _{c}$, which is small
compared with $\Delta $, Fig.~\ref{fig3}. When $\Gamma \ll \Delta ,$ one
can replace the last $\arctan $ in Eq.(\ref{n4}) by $\pi /2$ and resolve
this transcendental equation with respect to the bias voltage as $%
eV/2\approx \Delta +2\tilde{U}n-\Gamma \cot (2\pi n)$. The boundary of
bi-stability is found from $dV/dn=0$ , which yields the terminal value of
the level width $\Gamma _{c}=|\tilde{U}|/\pi $. The dot population and the
current, found from Eq.(\ref{eq:Is}) using Eqs.(\ref{dos4},\ref{n4}) remain
double-valued in the voltage region of bi-stability, Fig.~\ref{fig3}, where
both high-current and low-current branches are stable, while the
intermediate branch with $dn/dV<0$, Fig.~\ref{fig3}$a$, is unstable \cite%
{alebrastan}. Importantly, the coupling narrows the voltage range of the
hysteresis loop, but the transition from the low (high)-current branch to
the high (low)-current branch remains discontinuous as long as $\Gamma
<\Gamma _{c}$. The Hartree-type approximation of the correlation potential
is fully justified by the exact solution of Ref. \cite{alebrastan} in the
weak molecular-lead coupling regime.

The actual mechanisms of nanocircuit switching are of the highest
experimental and theoretical value. Further progress will depend upon
understanding of \emph{intrinsic} mechanisms of nano-memristors. Here, we
have extended our multi-polaron theory of the \emph{current controlled}
switching mechanism to the adiabatic molecular-size bridges with a
significant coupling to the leads ($\Gamma \gg \omega _{0}$). The degenerate
(or multilevel) MQD shows the \emph{hysteretic memory}, if the degeneracy of
the level (or the number of levels) is large enough, the Coulomb repulsion
is over-screened by EVI or by any other attractive mechanism, and the
coupling to the leads is below the critical value, Fig.~\ref{fig3}.
Importantly, the switching \emph{does not exist} in \emph{non-}degenerate
molecular dots neither weakly, Fig.~1, nor strongly coupled to the leads,
Fig.~\ref{fig2}b, contrary to the claims in Refs.~\cite{rat2,gal,gal2}. In
all cases, bi-stability is destroyed by a very strong coupling to the leads,
$\Gamma >\Gamma _{c}$.

ASA was supported by EPSRC (UK) (grant no. EP/H004483).

\end{document}